\documentclass[iop, twocolumn]{aastex631}%linenumbers,
\hypersetup{citecolor=blue,urlcolor=red}
\usepackage{multirow}
\usepackage{amsmath}
\usepackage{graphicx}
\usepackage{subfigure}
\usepackage{hyperref}  
\usepackage{nameref}
\usepackage{threeparttable}

\shorttitle{PSR~J0002$+$6216}
\shortauthors{Wei et al.}

\begin{document}

\title{The Timing and Polarization of PSR~J0002+6216}

\correspondingauthor{N. Wang, J. M. Yao}
\email{na.wang@xao.ac.cn, yaojumei@xao.ac.cn}

\author{Y. Wei\href{https://orcid.org/0009-0008-4753-5666}}
\affiliation{School of Physical Science and Technology, Xinjiang University,Urumqi, Xinjiang, 830046, People's Republic of China}
\affiliation{Xinjiang Astronomical Observatory, Chinese Academy of Sciences, 150 Science 1-Street, Urumqi, Xinjiang 830011, People's Republic of China; \textcolor{blue}{na.wang@xao.ac.cn, yaojumei@xao.ac.cn}}
\affiliation{Key Laboratory of~radio Astronomy, Chinese Academy of Sciences, Urumqi, Xinjiang, 830011, People's Republic of China}
\affiliation{Xinjiang Key Laboratory of~radio Astrophysics, 150 Science1-Street, Urumqi, Xinjiang, 830011, People's Republic of China}

\author{N. Wang\href{https://orcid.org/0000-0002-9786-8548}}%{\includegraphics[scale=0.04]}} %\thanks{E-mail: na.wang@xao.ac.cn}
\affiliation{Xinjiang Astronomical Observatory, Chinese Academy of Sciences, 150 Science 1-Street, Urumqi, Xinjiang 830011, People's Republic of China; \textcolor{blue}{na.wang@xao.ac.cn, yaojumei@xao.ac.cn}}
\affiliation{Key Laboratory of~radio Astronomy, Chinese Academy of Sciences, Urumqi, Xinjiang, 830011, People's Republic of China}
\affiliation{Xinjiang Key Laboratory of~radio Astrophysics, 150 Science1-Street, Urumqi, Xinjiang, 830011, People's Republic of China}

\author{J. P. Yuan\href{https://orcid.org/0000-0002-5381-6498}}%{\includegraphics[scale=0.04]}
\affiliation{Xinjiang Astronomical Observatory, Chinese Academy of Sciences, 150 Science 1-Street, Urumqi, Xinjiang 830011, People's Republic of China; \textcolor{blue}{na.wang@xao.ac.cn, yaojumei@xao.ac.cn}}
\affiliation{Key Laboratory of~radio Astronomy, Chinese Academy of Sciences, Urumqi, Xinjiang, 830011, People's Republic of China}
\affiliation{Xinjiang Key Laboratory of~radio Astrophysics, 150 Science1-Street, Urumqi, Xinjiang, 830011, People's Republic of China}
  
\author{J. M. Yao\href{https://orcid.org/0000-0002-4997-045X}}
\affiliation{Xinjiang Astronomical Observatory, Chinese Academy of Sciences, 150 Science 1-Street, Urumqi, Xinjiang 830011, People's Republic of China; \textcolor{blue}{na.wang@xao.ac.cn, yaojumei@xao.ac.cn}}
\affiliation{Key Laboratory of~radio Astronomy, Chinese Academy of Sciences, Urumqi, Xinjiang, 830011, People's Republic of China}
\affiliation{Xinjiang Key Laboratory of~radio Astrophysics, 150 Science1-Street, Urumqi, Xinjiang, 830011, People's Republic of China}

\author{M. Y. Ge}
\affiliation{Key Laboratory of Particle Astrophysics, Institute of High Energy Physics, Chinese Academy of Sciences, Beijing 100049, People's Republic of China}

\author{S. J. Dang\href{https://orcid.org/0000-0002-2060-5539}}
\affiliation{School of Physics and Electronic Science, Guizhou Normal University, Guiyang 550001, People's Republic of China}
\affiliation{Guizhou Provincial Key Laboratory of Radio Astronomy and Data Processing, Guizhou Normal University, Guiyang 550001, People's Republic of China}

\author{D. Zhao\href{https://orcid.org/0009-0007-8062-1454}}
\affiliation{Xinjiang Astronomical Observatory, Chinese Academy of Sciences, 150 Science 1-Street, Urumqi, Xinjiang 830011, People's Republic of China; \textcolor{blue}{na.wang@xao.ac.cn, yaojumei@xao.ac.cn}}
\affiliation{Key Laboratory of~radio Astronomy, Chinese Academy of Sciences, Urumqi, Xinjiang, 830011, People's Republic of China}
\affiliation{Xinjiang Key Laboratory of~radio Astrophysics, 150 Science1-Street, Urumqi, Xinjiang, 830011, People's Republic of China}

\author{F. F. Kou}
\affiliation{Xinjiang Astronomical Observatory, Chinese Academy of Sciences, 150 Science 1-Street, Urumqi, Xinjiang 830011, People's Republic of China; \textcolor{blue}{na.wang@xao.ac.cn, yaojumei@xao.ac.cn}}
\affiliation{Key Laboratory of~radio Astronomy, Chinese Academy of Sciences, Urumqi, Xinjiang, 830011, People's Republic of China}
\affiliation{Xinjiang Key Laboratory of~radio Astrophysics, 150 Science1-Street, Urumqi, Xinjiang, 830011, People's Republic of China}

\author{P. Liu}
\affiliation{Department of Astronomy, Xiamen University, Xiamen, Fujian 361005, People's Republic of China}

\author{J. T. Bai}
\affiliation{Xinjiang Astronomical Observatory, Chinese Academy of Sciences, 150 Science 1-Street, Urumqi, Xinjiang 830011, People's Republic of China; \textcolor{blue}{na.wang@xao.ac.cn, yaojumei@xao.ac.cn}}
\affiliation{Key Laboratory of~radio Astronomy, Chinese Academy of Sciences, Urumqi, Xinjiang, 830011, People's Republic of China}
\affiliation{Xinjiang Key Laboratory of~radio Astrophysics, 150 Science1-Street, Urumqi, Xinjiang, 830011, People's Republic of China}

\begin{abstract}
The combined timing analysis of data from the Five-hundred-meter Aperture Spherical Radio Telescope (FAST) and the Fermi Large Area Telescope (Fermi-LAT) confirmed that PSR~J0002+6216 is not a hyper-velocity (exceeding 1000~km~s$^{-1}$) pulsar. From this analysis, we determined the total proper motion of PSR~J0002+6216 to be $\mu_{\rm tot}=39.05\pm15.79$~mas~yr$^{-1}$, which is consistent with Very Long Baseline Interferometry (VLBI) measurements to within 0.24$\sigma$. Moreover, two glitches were detected for the first time, which occurred on MJD 58850(17) and MJD 60421(6), respectively. The second glitch exhibited an exponential recovery process, with $Q = 0.0090(3)$ and $\tau_{\rm d} = 45(3)$ days. Additionally, with FAST high-sensitivity observations, we measured the interstellar rotation measure (RM) and the three-dimensional (3D) orientation of the spin axis for the first time, and updated the dispersion measure (DM) of PSR~J0002+6216. Currently, no variations in RM or DM have been detected. By combining the measured RM with the observed position angle of the spin axis, we determined that the intrinsic position angle of the pulsar's spin axis is $\psi_{0}(\rm intrinsic)=89\fdg9\pm4\fdg6$. When we compared this with the proper motion position angle obtained from VLBI, we found a misalignment of approximately 23$^{\circ}$ between the spin and velocity angles of PSR~J0002+6216. At present, pulsars with 2D spin-velocity angle measurements are unable to fully test the \citet{2022ApJ...926....9J} model. However, with more high-precision observational data in the future, we will be able to further test models related to pulsar birth. 
\end{abstract}


%%%%%%%%%%%%%%%%%%%%%%%%%%%%%%%%%%%%%%%%%%%%%%%%%%
\section{Introduction}
%%%%%%%%%%%%%%%%%%%%%%%%%%%%%%%%%%%%%%%%%%%%%%%%%%
Measuring the proper motion of pulsars is crucial for understanding their velocity distribution and the origin of their velocities \citep{2005MNRAS.360..974H}. For young pulsars, it is particularly important for confirming their association with supernova remnants (SNR) and studying their spin-velocity relationship \citep{2021NatAs...5..788Y,2022ApJ...939...75Y}. Very Long Baseline Interferometry (VLBI) and timing are generally reliable methods for measuring pulsar proper motion. However, for certain pulsars, the results obtained from these two techniques can be inconsistent. For PSR~J0437$-$4715, \cite{2008ApJ...685L..67D} found that the proper motion value obtained from VLBI measurements differed by 4$\sigma$ in both right ascension and declination compared to the values derived from timing. This discrepancy was likely due to small shifts in the centroid position of the phase reference source used in the VLBI observations. Unmodeled timing noise and an imperfect timing model can lead to incorrect proper motion measurements, resulting in discrepancies between timing-based and VLBI-derived proper motion values. For example, \citet{2016ApJ...828....8D} showed that the proper motion measurements for PSR~J1022+1001 and PSR~J2145$-$0750, obtained from timing, are inaccurate, with discrepancies of up to 5$\sigma$ when compared to the more precise VLBI values. Given the low ecliptic latitudes of these pulsars, this discrepancy may stem from an imperfect model of the solar wind. In another case, \citet{2018ApJ...862..139D} found that the proper motion of PSR~B1913+16, as measured by VLBI, differed by more than 4$\sigma$ from the timing values reported by \citet{2016ApJ...829...55W}. This discrepancy is thought to be due to timing noise or variations in the dispersion measure (DM) within the timing data set for PSR~B1913+16. Considering that both timing and VLBI methods have their own biases, it is beneficial to measure a pulsar's proper motion using both techniques. 

Compared to the pulsars mentioned above, the proper motion values obtained from timing and VLBI for the young pulsar PSR~J0002+6216 show a much larger discrepancy. PSR~J0002+6216 is an interesting gamma-ray pulsar first discovered in Fermi Large Area Telescope (LAT) data \citep{2017ApJ...834..106C}. It is an isolated pulsar with  period of 115~ms and period derivative of $5.97\times10^{-15}~\rm s\ s^{-1}$. Subsequently, faint~radio pulsations were detected at Effelsberg \citep{2018ApJ...854...99W}. The pulsar is located near the SNR CTB~1, at least in the sky plane. A physical association was suggested by the discovery of a faint optical jet, apparently a pulsar bow-shock wind nebula, that points directly at the geometric centre of CTB~1 \citep[][henceforth Paper I]{2019ApJ...876L..17S}. The estimated distance to CTB~1 is $2.0\pm0.4$~kpc \citep{1982AJ.....87.1379L}, and it's mean measured age is $10\pm2$~kyr \citep{1994ApJ...434..635H,1997ApJ...488..307C,2006ApJ...647..350L}. An association between the pulsar and CTB~1 suggests that both originated from the same event, indicating that the pulsar is much younger than its characteristic age of $3\times10^6$~yr, consistent with its high-energy pulsations.  Paper I also carried out a timing analysis of the Fermi-LAT data from August 2008 to November 2018, deriving a relatively large proper motion, about 115~mas~yr$^{-1}$, albeit with a 30\% uncertainty. These parameters suggest that PSR~J0002+6216 is a so-called ``hyper-velocity" pulsar ($V_\text{PSR}>$1000 km\,s$^{-1}$), with an estimated transverse velocity of $1600\pm450$~km~s$^{-1}$, about four times the mean pulsar transverse velocity of 400~km~s$^{-1}$ \citep{2005MNRAS.360..974H}. More recently, using three years of VLBI observations from the High Sensitivity Array, \citet[][henceforth Paper II]{2023ApJ...958..163B} updated the total proper motion to $\mu$$_{\rm tot}=35.30\pm0.60$~mas~yr$^{-1}$ and the transverse velocity to $335\pm6$~km~s$^{-1}$, which is only a quarter of the estimates derived from Fermi-LAT timing. If PSR~J0002+6216 originated from the geometric center of CTB~1, the kinematic age would increase from $10.0\pm0.2$~kyr to $47.6\pm0.8$~kyr, based on the angular offset between PSR~J0002+6216 and CTB~1, as well as the newly measured proper motion. Two entirely different results for the proper motion of PSR J0002+6216 enlighten us to further investigate the pulsar. 
%Now, there are two entirely different results for the proper motion of PSR~J0002+6216. 

 As early as 1975, \citet{1975Natur.254..676T} discovered through calculations that the~radiation from an off-axis magnetic dipole would cause the spin-velocity alignment of pulsars, which was known as the electromagnetic rocket model. In observations over the past few decades, many young pulsars have been found to exhibit a 2D spin-velocity alignment, where the projection of the spin axis onto the sky is closely aligned with the direction of the pulsar's proper motion \citep[e.g.,][]{2001ApJ...556..380H,2012MNRAS.423.2736N,2013MNRAS.430.2281N}. More recently, with high-sensitivity observations from FAST, \citet{2021NatAs...5..788Y} found the first evidence of 3D spin-velocity alignment in PSR~J0538+2817, which is associated with the SNR~S147. Young pulsars, particularly those associated with SNR, are key objects for studying the spin-velocity relationship. This can help reveal the mechanisms that pulsars undergo during their formation. Therefore, for PSR~J0002+6216, it is crucial to conduct timing and polarization studies to verify the proper motion inconsistency, to determine the spin-velocity angle, and to explore the connection between PSR~J0002+6216 and CTB~1. 

%%% *** Timing + Polarization *** 
Pulsar timing can be carried out by comparing the Times of Arrival (ToAs) obtained from observations with those predicted by models, which results in timing residuals. With the increasing integration of observations across various frequencies, multi-band timing has become an important tool for pulsar timing studies \citep{2017MNRAS.466.1234Y, 2024arXiv240815022L}. Long-term timing observations reveal irregularities in pulsar timing, primarily due to timing noise and glitches \citep{2010MNRAS.402.1027H}. Glitches are sudden accelerations in the rotation of pulsars, which are rare and irregular phenomena, covering a range of sizes from $\sim 10^{-10}$ to $\sim 10^{-5}$ according to the value of $\Delta\nu/\nu$ \citep{2013MNRAS.429..688Y}. Glitches are thought to be caused by an abrupt transfer of angular momentum from the interior superfluid to the crust of pulsars \citep{2013PhRvL.110a1101C} or by crustquakes in neutron stars \citep{2020ApJ...897..173B}. Since Effelsberg detected faint radio pulses from PSR~J0002+6216, it has only obtained the total intensity pulse profile, dispersion measure (DM), and other parameters. However, polarimetry measurements were not possible \citep{2018ApJ...854...99W}. As demonstrated in \cite{2022ApJ...939...75Y, 2021NatAs...5..788Y}, high-sensitivity polarization observations with FAST can provide measurements of the RM and also enable the determination of the 3D spin axis direction through a Rotational Vector Model (RVM) fit. In this paper, by combining Fermi LAT and FAST observations, we detected glitches in PSR~J0002+6216 for the first time, and measured its proper motion. We also obtained the Rotation Measure (RM), updated its DM, and calculated the 2D spin-velocity angle for this pulsar. 

The paper is structured as follows: in Section~\ref{sec:2}, we introduced the Fermi-LAT and FAST observations and data processing; in Section~\ref{sec:3}, we present the timing and polarization results and discuss the 2D spin-velocity relationship; in Section~\ref{sec:4}, we discuss the results and summarize the paper.

%%%%%%%%%%%%%%%%%%%%%%%%%%%%%%%%%%%%%%%%%%%%%%%%%%
\section{Observations and Data Processing} \label{sec:2}
%%%%%%%%%%%%%%%%%%%%%%%%%%%%%%%%%%%%%%%%%%%%%%%%%%

%%%%%%%%%%%%%%%%%%%%%%%%%%%%%%%%%%%%%%%%%%%%%%%%%%
\subsection{Fermi-LAT Data} \label{sec:Fermi Data}
%%%%%%%%%%%%%%%%%%%%%%%%%%%%%%%%%%%%%%%%%%%%%%%%%%
The launch of the Fermi satellite in 2008, equipped with  the LAT, has significantly advanced the study of gamma-ray emissions from pulsars. The LAT, with its large effective area and wide field of view, provides highly sensitive observational data ranging from 20~MeV to approximately 300~GeV \citep{2014ARA&A..52..211C}. In this paper,  we present a comprehensive analysis of timing data from PSR~J0002+6216, derived using the Fermi Science Tools (v11r5p3)\footnote{\url{https://fermi.gsfc.nasa.gov/ssc/data/analysis/scitools/}} from nearly 16 years of Fermi-LAT observations collected between September 2008 and March 2024. The events were constructed using {\tt gtselect} with an angular distance of less than $0\fdg5$, a zenith angle of less than 105$^{\circ}$, and an energy range of 0.1 -- 10~GeV \citep{2011ApJS..194...17R}. Then, following \cite{2014ascl.soft12012K}, {\tt GeoTOA} software package was used to generate the Geocentric ToAs. Each ToA was derived from observational data accumulated over an exposure period of either 32 or 69 days, following the procedure described by \cite{2019NatAs...3.1122G, 2020ApJ...900L...7G}; \cite{2024ApJ...977..243Z}.

%%%%%%%%%%%%%%%%%%%%%%%%%%%%%%%%%%%%%%%%%%%%%%%%%%
\subsection{FAST Data} \label{sec:FAST Data}
%%%%%%%%%%%%%%%%%%%%%%%%%%%%%%%%%%%%%%%%%%%%%%%%%%
From 2021 to 2025, we conducted 45 observations of PSR~J0002+6216 using the central beam of the 19-beam receiver of the FAST telescope. Most of these observations were performed monthly, with each session lasting half an hour, except for three observations made on MJDs 59456, 60024 and 60189, which lasted two hours. The FAST 19-beam receiver covers the frequency range from 1000~MHz to 1500~MHz. Due to reduced sensitivity at the edges of the band, the effective bandwidth is 400~MHz, spanning from 1050~MHz to 1450~MHz. We recorded all data in PSRFITS format,  incorporating four polarizations, 4096 frequency channels, and a sampling time of 49.152~$\mu$s. At the beginning of each observation, a 100~s calibration noise diode is employed for polarization calibration. Since the 1150~MHz to 1300~MHz band experiences more radio-frequency interference (RFI) than the other bands, after getting rid of RFI, we conducted timing analysis using data from 1050~MHz to 1450~MHz, while we conducted polarization analysis  using the relatively continuous data from 1300~MHz to 1450~MHz. 

We processed our data using the {\tt DSPSR} analysis program \citep{DSPSR} and the {\tt PSRCHIVE} software package \citep{PSRCHIVE}. First, we folded the data for each channel using the {\tt DSPSR} based on the rotational ephemeris provided by \cite{2017ApJ...834..106C}. Then, after removing the RFI with {\tt pazi}, we performed polarization calibration on the data. 

For timing analysis, we first summed the data across time, polarization, and frequency to produce mean pulse profiles. The phases of these profiles were then aligned by using {\tt pas}, and {\tt psradd} was used to combine them into a high signal-to-noise ratio (S/N) standard pulse profile. Next, the pulse ToAs relative to FAST site were determined by cross-correlating the standard template with the observed profiles using the {\tt pat} tool. Then, {\tt TEMPO2} was used to correct both Fermi and FAST ToAs to the the Solar system with DE438 and Barycentric Dynamical Time (TDB). Finally, TEMPO2 was employed to fit a pulsar timing model to the corrected ToAs, yielding a timing solution \citep{2006MNRAS.372.1549E}. The pulse phase $\phi$ in the standard timing model is given by:
\begin{equation}
\label{equ:1}
\phi(t)=\phi_{0}+\nu(t-t_{0})+\frac{\dot\nu}{2}(t-t_{0})^{2}+\frac{\ddot\nu}{6}(t-t_{0})^{3}+\cdots
\end{equation}
where $\phi_{0}$ is the pulse phase at the fiducial epoch $t_{0}$, $\nu$, $\dot\nu$, and $\ddot\nu$ represent the spin frequency and the first and second derivatives of the spin frequency, respectively.

%%%%%%%%%% table 1 %%%%%%%%%%
\begin{table*}
\caption{PSR J0002$+$6216's pre- and post-glitch timing solutions.}
\label{tab:Glitch1}
\renewcommand{\arraystretch}{1.25}
    \centering
    \begin{tabular}{ccccccccc}
    \hline
    \hline       
    Inter-glitch & Epoch  & $\nu$  & $\dot{\nu}$ & $\ddot{\nu}$  & $N_{\rm ToA}$  & MJD Range   & RMS & Telescope \\
     & (MJD)  & (Hz)   & ($10^{-13}$ s$^{-2}$)   &  ($10^{-24}$ s$^{-3}$) &  & (MJD) & ($\mu$s) \\
    \hline     
    Pre-G1 &56775 &8.6682102904(1) &$-$4.48318(1) &28(4)  &51  &54734 -- 58816  &4601&Fermi-LAT   \\  
    G1-G2 &59641&8.6680992984(3) &$-$4.48312(7) &$-$36(60)  &39  &58885 -- 60397  &3183 &Fermi-LAT  \\ 
    G1-G2 &59826 &8.66809213278(7) &$-$4.48302(2) &$-$69(24) &25  &59244 -- 60408  &424 &FAST  \\
    Post-G2 &60563 & 8.6680845708(9) &$-$4.581(2) &679(65)  &20  &60433 -- 60693  &1004 &FAST   \\   
    \hline
    \end{tabular}
%\end{center} \vspace{-0.4cm}
\end{table*}
%%%%%%%%%% table 1 %%%%%%%%%%

In order to determine the accurate position, we use the {\tt SPECTRALMODEL} plugin from the {\tt TEMPO2} \citep{TEMPO2} software package to estimate the red noise model in the timing residuals, as described by \citet{2020ApJ...896..140D}. The power law model of red noise is characterized by three parameters: amplitude ($A$), spectral index ($\alpha$), and corner frequency ($f_{c}$). The expression of the model is as follows: $P(f)=A/[1+(f/f_{c})^{2}]^{\alpha/2}$. Based on this model, we adopted the iterative process described by \citet{2011MNRAS.418..561C}, which allowed us to obtain unbiased proper motion measurements, even in the presence of pulsar timing noise. Finally, we conducted a fitting analysis on Fermi-LAT and FAST data by using both the classic pulsar timing software package {\tt TEMPO2} and the modern pulsar timing software package {\tt PINT} \citep{PINT} \footnote{\url{https://github.com/nanograv/PINT}}.

For polarization analysis, we first used {\tt paz} to extract the data from 1300~MHz to 1450~MHz; Next, we utilized the {\tt rmfit} tool to obtain the observed rotation measure (RM$_{\rm obs}$) at frequencies centered at 1375~MHz. Then, we used {\tt psredit} to revise the RM value in the header file. Finally, we did the RVM-fit by using the {\tt psrmodel}, and obtained the direction of 3D spin axis.

%%%%%%%%%%%%%%%%%%%%%%%%%%%%%%%%%%%%%%%%%%%%%%%%%%
\section{Results} \label{sec:3}
%%%%%%%%%%%%%%%%%%%%%%%%%%%%%%%%%%%%%%%%%%%%%%%%%%

In this section, we present the timing results from Fermi-LAT and FAST observations, as well as the polarization results from FAST. Through timing analysis, we detected two glitches in the young pulsar PSR~J0002+6216 for the first time. After excluding  the ToAs from the exponential recovery process,  we obtained the proper motion measurement. In the polarization analysis, we reported the first measurement of the RM and determined the 3D orientation of the spin axis by obtaining the position angle of the spin axis ($\psi_0$) and the inclination angle ($\zeta$) relative to our line of sight. Finally, we identified a significant 2D misalignment between the polarization position angle (PPA) and the proper motion vector of PSR~J0002+6216.

%%%%%%%%%%%%%%%%%%%%%%%%%%%%%%%%%%%%%%%%%%%%%%%%%%
\subsection{Glitch} \label{sec:Glitch}
%%%%%%%%%%%%%%%%%%%%%%%%%%%%%%%%%%%%%%%%%%%%%%%%%%
%%%%%%%%%% table 2 %%%%%%%%%%
\begin{table}[htb]
\caption{The fitted timing solutions and glitch parameters for PSR J0002+6216.}
\label{tab:Glitch2}
\renewcommand{\arraystretch}{1.25}
\centering
\begin{tabular}{lcc}
\hline
\hline  
Parameter  &Glitch 1 &Glitch 2  \\
\hline
$\nu$ (Hz)  &8.6681396800(7)   &8.66808248785(5) \\
$\dot{\nu}$ ($10^{-13}$\rm\ s$^{-2}$) &$-$4.4845(5)   & $-$4.48305(4)  \\
Freq. epoch (MJD) &58598 & 60075 \\
Glitch epoch (MJD)  &58850(17)   &60421(6)  \\
Data range (MJD) &57571 -- 59626&59456 -- 60693  \\
TOA numbers &30 &44  \\
Rms residual ($\mu {\rm s}$) &2881 &269  \\
$\Delta\phi$  &0.05(4)  &0.30(1)  \\
$\Delta\nu$ ($10^{-9}$\rm\ Hz) &19(2) & 21243(7) \\
$\Delta\nu/\nu$ ($10^{-9}$)&2.2(3) &2450.7(8)  \\
$\Delta\dot{\nu}$ ($10^{-16}$\rm\ s$^{-2}$) &$-$5(1) &$-$559(37)  \\
$\Delta\dot{\nu}/\dot{\nu}$ ($10^{-3}$) &1.0(3) &125(8)  \\
$\tau_{\rm d}$ &--  &45(3)   \\
$Q$ &--  &0.0090(3)  \\
\hline
\end{tabular}
\end{table}
%%%%%%%%%% table 2 %%%%%%%%%%

By analyzing the timing residuals from Fermi-LAT and FAST data, we detected glitches in PSR~J0002+6216 for the first time. There are two glitches: one occurred at the end of the Fermi-LAT data on MJD~58850(17), and the other occurred at the end of the FAST data on MJD~60421(6). Table~\ref{tab:Glitch1} presents the timing solutions before and after the glitches, produced by {\tt TEMPO2}, along with the 1$\sigma$ uncertainties of the parameters. The rotation parameters ($\nu$, $\dot\nu$ and $\ddot{\nu}$) were obtained by fitting Equation~(\ref{equ:1}). As shown in Table~\ref{tab:Glitch2}, we listed the timing solutions along with the values and uncertainties of the glitch parameters that describe these two events. To determine the glitch parameters listed in Tables~\ref{tab:Glitch1} and~\ref{tab:Glitch2} from the timing fits, we held the position and proper motion fixed. The position parameters derived from Table~\ref{tab:Position} at epoch MJD 59867 were used, while the proper motion parameters were adopted from those reported in Paper II. The specific details are as follows:

Glitch events typically cause additional phase changes, which can be modeled by the following formula \citep{2006MNRAS.372.1549E, 2013MNRAS.429..688Y}:
\begin{equation}
\begin{split}
\label{equ:2} 
\phi_{\rm g} = \Delta\phi+ \Delta\nu_{\rm p}(t - t_{\rm g}) +   \frac{1} {2} \Delta\dot{\nu}_{\rm p} (t - t_{\rm g})^{2} \\
+  { [1-e^{-(t - t_{\rm g})/\tau_{\rm d}}] \Delta\nu_{\rm d}\tau_{\rm d} }  \ ,
\end{split}
\end{equation}
where $\Delta\phi$ is the offset of the pulse phase at the glitch epoch $t_{g}$, $\Delta\nu_{p}$ and $\Delta\dot\nu_{p}$ are the permanent increments in frequency and its first derivative, $\Delta\nu_{d}$ is the temporary frequency increment, and $\tau_{d}$ is the exponential decay timescale. Hence, the changes in frequency and its first derivative can be expressed as:
\begin{equation}
\label{equ:3} 
\frac{\Delta\nu_{\rm g}}{\nu}=\frac{\Delta\nu_{\rm p}+\Delta\nu_{\rm d}}{\nu},
\end{equation}

\begin{equation}
\label{equ:4} 
 \frac{\Delta\dot{\nu_{\rm g}}} {\dot{\nu}} = \frac{\Delta\dot{\nu}_{\rm p} - {\Delta\nu_{\rm d} / \tau_{\rm d}}} {\dot{\nu}} \ .
\end{equation}
Furthermore, the glitch recovery factor is defined as: $Q=\Delta\nu_{\rm d}/\Delta\nu_{\rm g}$.

However, since the glitch epoch could not be precisely determined from the observations, it was set at the midpoint between the last pre-glitch epoch and the first post-glitch epoch, with an uncertainty of one-quarter of the observation gap \citep{2011MNRAS.414.1679E}. Subsequently, the glitch parameters were determined by fitting Equation~(\ref{equ:2}), and their associated uncertainties were calculated using the standard error propagation formula.

\begin{figure}[!htpb]
\center
    \includegraphics[width=8.5 cm]{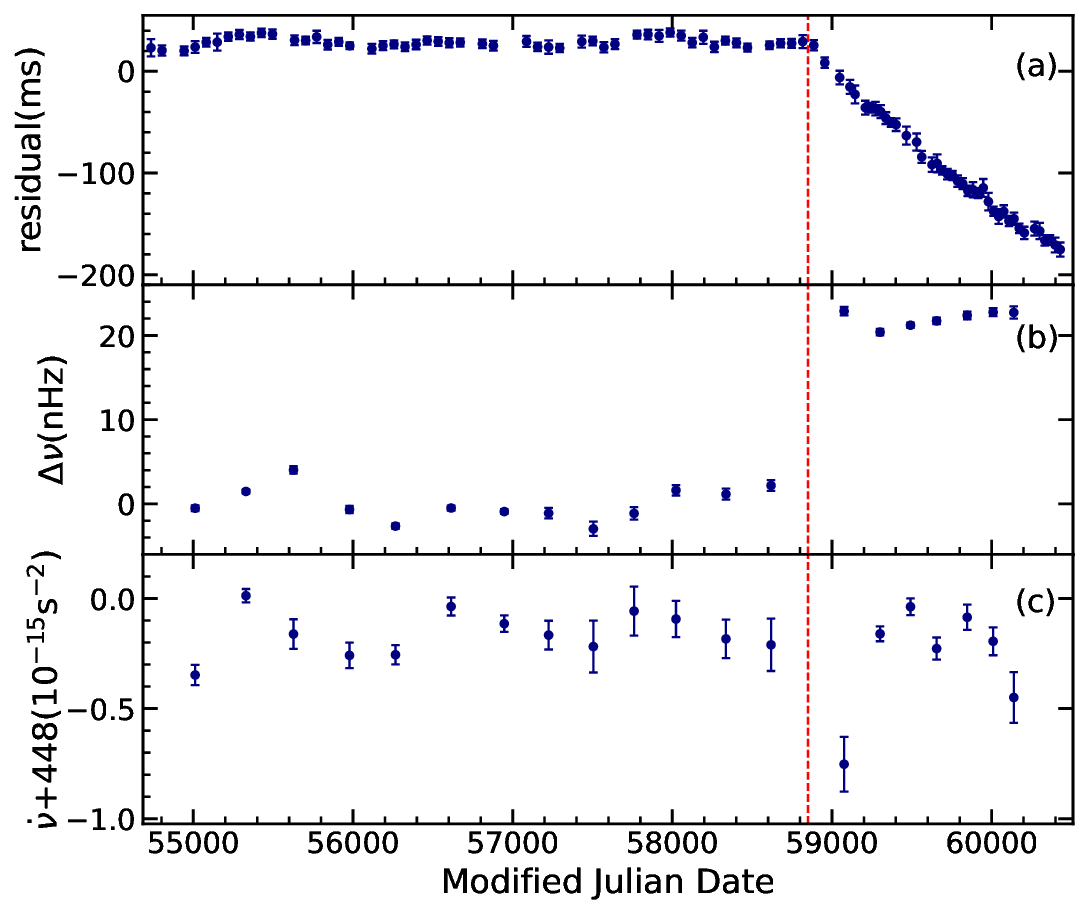}
    \caption{The observed first glitch for PSR~J0002+6216 from Fermi-LAT observations: (a) the timing residual relative to the pre-glitch spin-down model; (b) the relative change ($\Delta\nu$) of spin frequency, which is defined as the change in pulsar rotation frequency relative to the frequency model before the glitch; (c) the evolution of the spin-down rate ($\dot{\nu}$) over time before and after the glitch. The vertical line indicates the glitch epochs within our data span.}
    \label{fig:Fermi-glitch}
\end{figure}

The small amplitude glitch detected in the Fermi-LAT data is quite evident. Panel (a) of Figure~\ref{fig:Fermi-glitch} shows the timing residuals relative to the spin-down model prior to the glitch, which aligns with the typical pattern observed in small glitches. This is followed by a significant linear break on MJD~58850(17). A series of values of $\nu$ and $\dot{\nu}$ were obtained from independent fits of equation~(\ref{equ:1}) (omitting the $\ddot\nu$ term) to data spans ranging from 32 -- 69 d. The frequency residuals ($\Delta \nu$) are obtained by subtracting the pre-glitch model values from the observed values of spin frequencies $\nu$. Panels (b) and (c) reveal discontinuities in $\Delta \nu$ and the frequency derivative $\dot{\nu}$, with a fitted glitch size of $\Delta \nu/\nu \sim  2.2(3)\times10^{-9}$ and a slight decrease in the frequency derivative $\dot{\nu}$ of $-5(1)\times10^{-16}\rm \ s^{-2}$.

\begin{figure}[!htpb]
\center
    \includegraphics[width=8.5 cm]{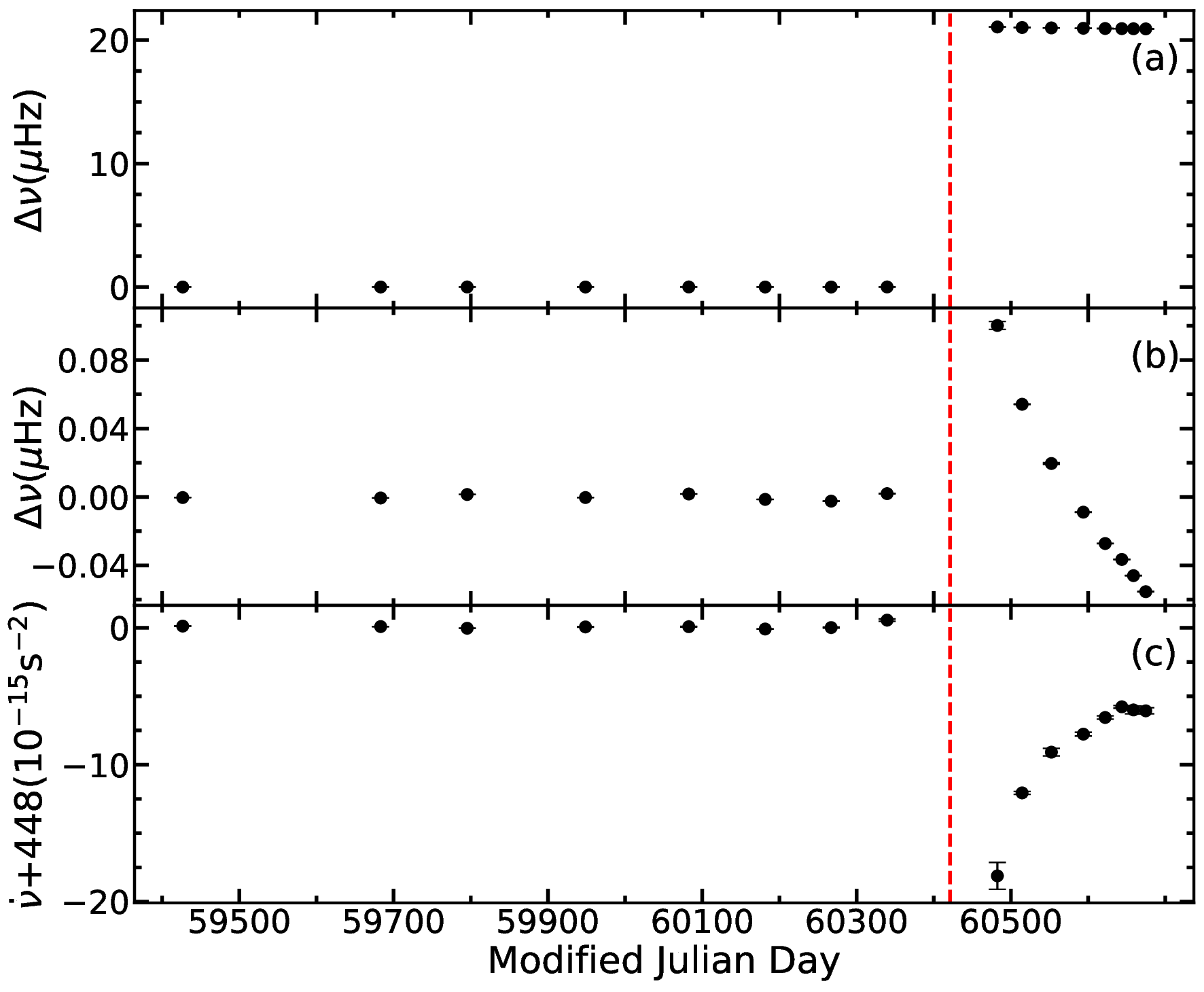}
    \caption{The observed second glitch for PSR~J0002+6216 from FAST observations: The vertical lines indicate the glitch epoch. Panel (a) displays the variations of spin frequency ($\Delta\nu$) relative to the pre-glitch spin-down model; Panel (b) is an expanded plot of $\Delta\nu$, showing the evolution details of $\Delta\nu$ after the glitch, obtained by subtracting its average value from $\Delta\nu$ of post-glitch; Panel (c) shows the variations of the first frequency derivative ($\dot{\nu}$)}.
    \label{fig:FAST-glitch}
\end{figure}

%%%%%%%%%% table 3 %%%%%%%%%%
\begin{table*}
\caption{The position of PSR J0002+6216 in J2000 Equatorial Coordinates measured through timing.}
\label{tab:Position}
\renewcommand{\arraystretch}{1.25}
    \centering
    \begin{tabular}{cccccccc}
    \hline
    \hline
    Right Ascension & Declination  & POSEPOCH & Data Range & Telescope &Software\\
    (h:m:s)	 & (d:m:s) & MJD & MJD & &\\
    \hline 
    00:02:58.1474(723) &$+$62:16:09.602(505) & 56498 &54734$-$58263	 &Fermi-LAT&TEMPO2 \\
    00:02:58.1487(723) &$+$62:16:09.598(505) & 56498 &54734$-$58263	 &Fermi-LAT&PINT \\
    \tableline
    00:02:58.2176(12) & $+$62:16:09.468(6) & 59867 &59510$-$60223 &FAST&TEMPO2 \\
    00:02:58.2176(12) & $+$62:16:09.468(6) & 59867 &59510$-$60223 &FAST&PINT \\
    \tableline
    00:02:58.2203(7) & $+$62:16:09.460(4) & 60071 &59822$-$60320 &FAST&TEMPO2  \\
    00:02:58.2203(7) & $+$62:16:09.460(4) & 60071 &59822$-$60320 &FAST&PINT  \\
    \hline
    \end{tabular}
\end{table*}
%%%%%%%%%% table 3 %%%%%%%%%%

\begin{figure*}[!htpb]
\center
    \includegraphics[width=16 cm]{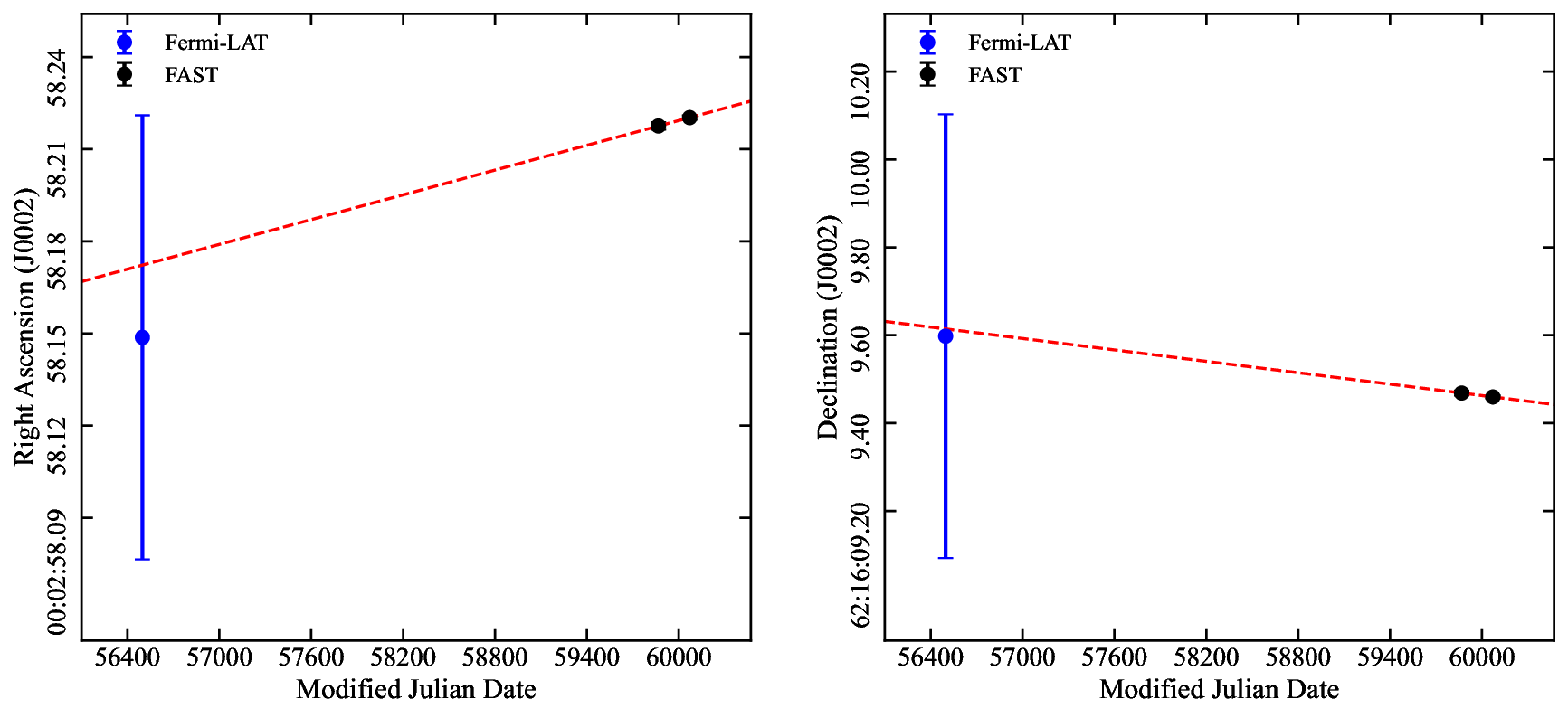}
    \caption{The positions in right ascension and declination for PSR~J0002+6216 versus time. The blue and black points represent Fermi data and FAST data, respectively. The red dashed lines show the best least square fit results for proper motion. The resulting proper motion values are listed in Table~\ref{tab:PM}.}
    \label{fig:PMRA-PMDec}
\end{figure*}

More than four years after the first small glitch, a larger glitch was detected on MJD~60421(6). The relative size of this glitch is $\Delta \nu/\nu \sim  2450.7(8)\times10^{-9}$ with a relative variation in the spin-down rate of $\Delta \dot{\nu}/\dot{\nu} \sim 125(8)\times10^{-3}$. The large glitch also resulted in a significant change in the evolution of the spin-down rate. Our analysis of the fitted timing residuals reveals a distinct exponential recovery component, as clearly illustrated in panel (b) of Figure~\ref{fig:FAST-glitch}. We incorporated the exponential decay term into the {\tt TEMPO2} fitting and determined an exponential recovery timescale of $45(3)$ days and a recovery factor of $Q= 0.0090(3)$. This indicates that the recovery process is relatively fast, as the coupling between the superfluid vortices and the inner crust of the pulsar, as described by the vortex creep model \citep{1989ApJ...346..823A}, facilitates the system's return to a new stable state over time. The vortex creep model effectively explains typical glitch behaviors and predicts the time of future glitches \citep{2022MNRAS.511..425G}.

%%%%%%%%%%%%%%%%%%%%%%%%%%%%%%%%%%%%%%%%%%%%%%%%%%
\subsection{Proper motion} \label{sec:Tim}
%%%%%%%%%%%%%%%%%%%%%%%%%%%%%%%%%%%%%%%%%%%%%%%%%%

Combining ToAs obtained from Fermi-LAT and FAST observations, we conducted a comprehensive timing analysis of the position, proper motion and spin evolution of PSR~J0002+6216. Table~\ref{tab:Position} displays the positions of PSR~J0002+6216, which were determined by utilizing pulsar timing software packages, namely {\tt TEMPO2} and {\tt PINT}. The first four columns list the Right Ascension (R.A.) and Declination (Dec.) in J2000 equatorial coordinates, along with their 1$\sigma$ uncertainties.  The subsequent two columns record the epochs for each position measurement and the time range of the corresponding data. As discussed in Section~\ref{sec:Glitch}, two glitches are observed at the end of both Fermi-LAT and FAST data. Due to the limited data available following each glitch, our determination of the position of PSR~J0002+6216 is solely based on the data collected prior to the occurrence of each glitch. Considering the higher measurement accuracy of the FAST telescope, we utilized observations from FAST across two different overlapping data spans to determine two positions for PSR~J0002+6216, where the red noise model is included and fixed. In comparison to two positions derived from FAST data, the position obtained from Fermi data exhibits significantly greater uncertainty. In Table~\ref{tab:Position}, the final column shows that each data sets was processed independently using {\tt TEMPO2} and {\tt PINT} to determine the positions and their associated uncertainties.

%%%%%%%%%% table 4 %%%%%%%%%%
\begin{table*}[htb]
\renewcommand{\arraystretch}{1.25}
\caption{ The proper motion and position angle of PSR~J0002+6216 derived from this work and from the referenced papers.}
\label{tab:PM}
    \centering
    \begin{threeparttable}
    \begin{tabular}{ccccc}
    \hline
    \hline
    $\mu_\alpha^*$ & $\mu_{\delta}$ & $\mu_{\text{tot}}$ &$\psi_{\rm pm}$  & Reference \\
    (mas yr$^{-1}$)	& (mas yr$^{-1}$) & (mas yr$^{-1}$) & (deg) & 	\\
    \hline 
 97 $\pm$ 32 &$-$57 $\pm$ 27& 115 $\pm$ 33&121 $\pm$ 13 &	Paper I \\
 32.52 $\pm$ 0.59& $-$13.71 $\pm$ 0.53 & 35.30 $\pm$ 0.60& 112.86 $\pm$ 0.83 &	Paper II \\
 35.74 $\pm$ 16.47&$-$15.73 $\pm$ 11.67 &39.05 $\pm$ 15.79 &113.76 $\pm$ 18.45 &	This Work  \\
    \hline
    \end{tabular}
    \begin{tablenotes}
    \item[] Notes. The proper motion in R.A. is $\mu_\alpha^*$ = $\mu_\alpha \cos \delta$ and in Dec. is $\mu_{\delta}$.
    \end{tablenotes}
    \end{threeparttable}
\end{table*}

%%%%%%%%%% table 4 %%%%%%%%%%

Given that the positions obtained using {\tt TEMPO2} and {\tt PINT} are identical, we rely on these to analyze the proper motion of PSR~J0002+6216. As shown in Figure~\ref{fig:PMRA-PMDec}, we utilized the curve\_fit function from Python's SciPy library \footnote{\url{https://docs.scipy.org/doc/scipy/reference/generated/scipy.optimize.curve_fit.html}} to perform a weighted least-squares fit, which allowed us to determine the slopes, that is, the proper motion in R.A. and in Dec. And the best-fit results are indicated by the red dashed lines in Figure~\ref{fig:PMRA-PMDec}. Furthermore, in Table~\ref{tab:PM}, we listed the best-fit values for the proper motion along with their corresponding uncertainties.

As presented in the last row of Table~\ref{tab:PM}, the proper motion in R.A. is $35.74\pm16.47$~mas~yr$^{-1}$ and in Dec. is $-15.73\pm 11.67$~mas~yr$^{-1}$. These values result in a total proper motion of $39.05\pm15.79$~mas~yr$^{-1}$, with a position angle of $113\fdg76 \pm 18\fdg45$, measured from north towards east. In the first two rows of Table~\ref{tab:PM}, we also listed the measured proper motions and their respective  position angles as reported in Paper I and Paper II. The newly measured proper motion in R.A. and Dec. are in agreement with VLBI measurements within a 0.24$\sigma$. This agreement enhances the reliability of our measurements and suggests that both the data and the methods used in both studies are reliable.  Additionally, when  comparing these results with the proper motion reported in Paper I, it is evident that the previous measurements derived from Fermi-LAT timing were unreliable. Based on the measurements from VLBI and timing in this paper, we confirmed that PSR~J0002+6216 is not a hyper-velocity pulsar. It is important to note that since the two positions obtained from FAST were derived from non-independent datasets, this may lead to an underestimation of the uncertainties in both position and proper motion. In the future, long-term, high-cadence FAST observations could provide more precise proper motion measurements for PSR J0002+6216 through pulsar timing.

%%%%%%%%%%%%%%%%%%%%%%%%%%%%%%%%%%%%%%%%%%%%%%%%%%
\subsection{Polarization of PSR~J0002+6216} \label{sec:Pol}

\subsubsection{Rotation measure results}\label{sec:RM}
\begin{figure}[!htpb]
\center
    \includegraphics[width=8.5 cm]{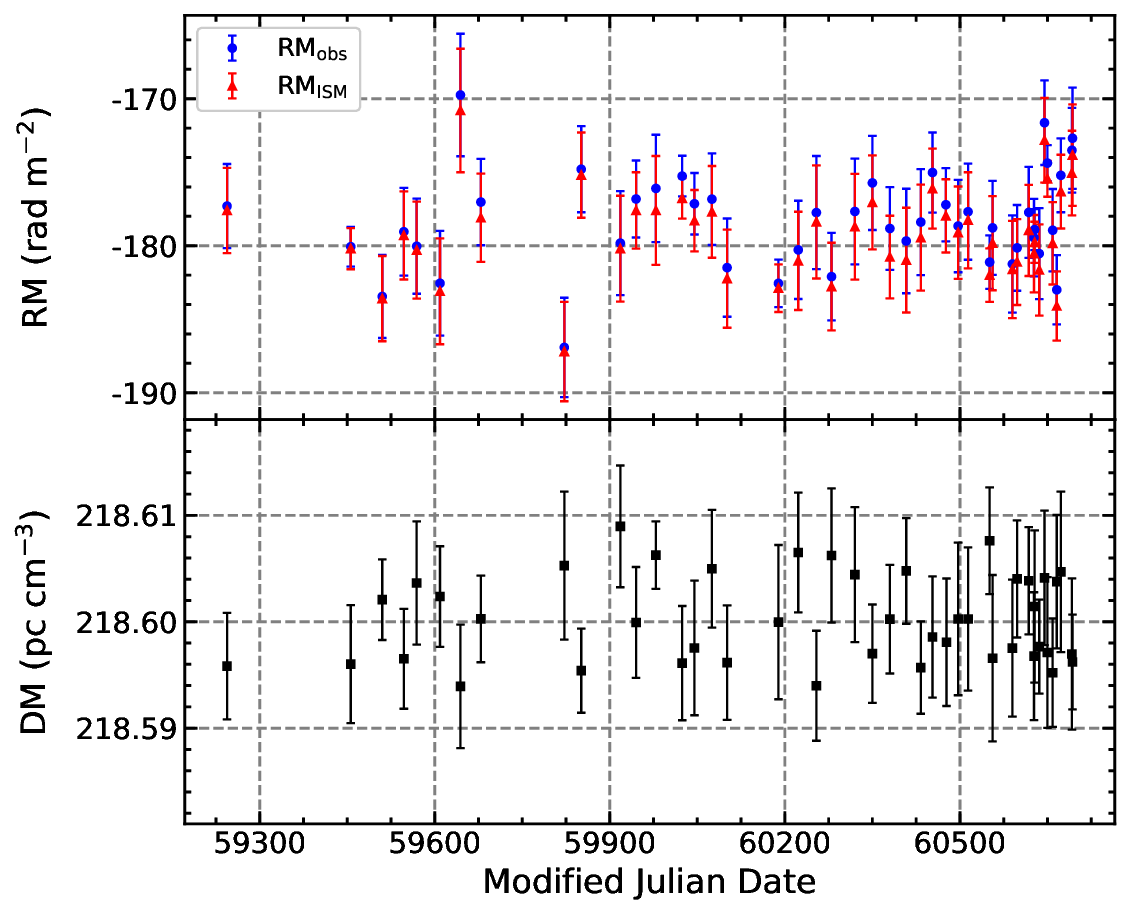}
    \caption{Variations of RM and DM for PSR~J0002+6216 as a function of the MJD. Upper panel: the blue and red data points represent the RM$_{\rm obs}$ and the RM$_{\rm ism}$, and the error bars show the 1$\sigma$ uncertainty interval. Lower panel: the black data points represent the measured DM.}
    \label{fig:RM-DM}
\end{figure}

For all these FAST observations, we used the {\tt RMFIT} to obtain the RM from data in the 1300~MHz to 1450~MHz range. For the pulsar, the RM$_{\rm obs}$ includes contributions from both the interstellar medium (RM$_{\rm ism}$) and the Earth's ionosphere (RM$_{\rm iono}$). To derive RM$_{\rm ism}$, we utilized the IONFR \citep{IONFR} and the values of the ionospheric electron column density from the NASA CDDIS GNSS website\footnote{\url{https://cddis.nasa.gov/archive/gnss/products/ionex/}} to estimate RM$_{\rm iono}$. We then calculated RM$_{\rm ism}$ using the relation $\rm RM_{\rm ism}=RM_{\rm obs}-RM_{\rm iono}$. In the upper panel of Figure~\ref{fig:RM-DM}, we show the variations of both RM$_{\rm obs}$ and RM$_{\rm ism}$ over time. The blue data points represent the RM$_{\rm obs}$ and the red data points represent RM$_{\rm ism}$, with the error bars indicating the 1$\sigma$ uncertainty associated with each measurement. For each observation, we measured the DM by using a standard timing analysis over the entire bandwidth of 1050 -- 1450~MHz. In the lower panel of Figure~\ref{fig:RM-DM}, we show the DM variations over time. 

%%%%%%%%%% figure RM %%%%%%%%%%
%%%%%%%%%% figure RM %%%%%%%%%%

From the derived RM$_{\rm ism}$ and the measured DM, we determined that the average values are $-179.161 \pm 2.973$~rad~m$^{-2}$ for RM$_{\rm ism}$  and $218.600 \pm 0.006$~pc~cm$^{-3}$ for DM. The corresponding standard deviations are 3.082~rad~m$^{-2}$ and 0.004~pc~cm$^{-3}$, respectively. From these results, we have not detected any variations in RM$_{\rm ism}$ and DM for the pulsar PSR J0002+6216, as their standard deviations are comparable to the uncertainties of their respective average values.

In addition, by using the average RM$_{\rm ism}$ value along with the average DM value, we estimated the mean magnetic field component parallel to the line of sight with the equation below:
\begin{equation}
\label{eq:RM}
\left<B_{\parallel}\right>=1.232\,\mu \text{G}\left(\frac{\text{RM}}{\text{rad}\, \text{m}^{-2}}\right)\left(\frac{\text{DM}}{\text{pc}\,\text{cm}^{-3}}\right)^{-1}.
\end{equation}
We derived the magnetic field strength is $-1.01\pm0.02$~$\mu\rm G$. PSR~J0002+6216 is a pulsar located within the Galactic disk and has a Galactic latitude (Gb) of $-$0.07~deg. From \cite{2018ApJS..234...11H}, the typical strength of the magnetic field in the Galactic disk is on the order of a few $\mu\rm G$, which means our measurement is in agreement with this.

%%%%%%%%%%%%%%%%%%%%%%%%%%%%%%%%%%%%%%%%%%%%%%%%%%
\subsubsection{The 3D spin axis} \label{sec:RVM}
\begin{figure}[!ht]
\center
    \includegraphics[width=12.0 cm, angle=270]{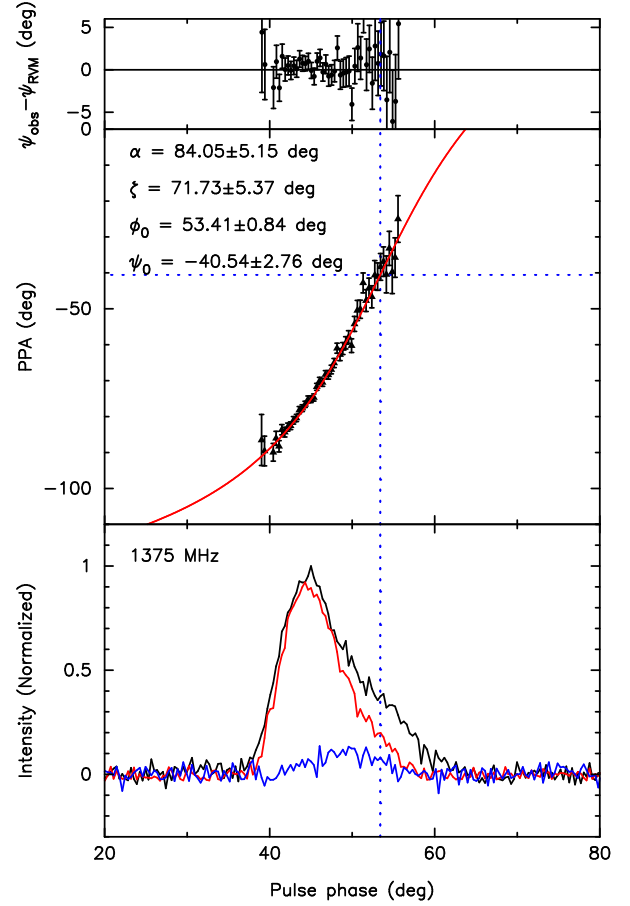}
    \caption{The polarization profile for PSR~J0002+6216 at 1375~MHz from observation made on MJD~59456. Bottom panel: the total intensity, the linear polarization and the circular polarization are represented by black, red, and blue lines, respectively. Middle panel: the observed PPAs ($\psi$) are shown as a function of pulse phase with the red curve indicating the best-fit RVM solution. Top panel: the fit residuals are shown as a function of pulse phase. The vertical and horizontal blue dotted lines represent the central pulse phase fitted by RVM and the corresponding PPA, respectively.}
    \label{fig:59456}
\end{figure}

%%%%%%%%%%%%%%%%%%%%%%%%%%%%%%%%%%%%%%%%%%%%%%%%%%
The duration of the observations conducted on MJDs 59456, 60024, and 60189 are 2 hours. Among these observations, the one conducted on MJD 59456 has the highest S/N. Therefore, we utilized this data for the subsequent RVM-fit to determine the 3D spin axis. As shown in Figure~\ref{fig:59456}, the polarization position angle (PPA) of PSR~J0002+6216 exhibits an S-shaped curve. Following the methodology of \cite{2022ApJ...939...75Y}, we performed an RVM-fit using the {\tt psrmodel}, with the red line representing the best-fit results. From this fit, we obtained the following parameters: $\alpha = 84\fdg05\pm5\fdg15$, $\zeta = 71\fdg73\pm5\fdg37$, $\phi_0 = 53\fdg41\pm0\fdg84$ and $\psi_0=-40\fdg54\pm2\fdg76$, where $\alpha$ is the angle between the spin axis and the magnetic axis, $\zeta$ denotes the inclination angle of the spin axis from the line of sight, and $\phi_0$ indicates the pulse phase for the closest approach of the line of sight to the magnetic axis, with a corresponding position angle (PA) of $\psi_0$. To compare $\psi_0$ with the position angle of pulsar velocity, it is necessary to adjust it for an infinite frequency, yielding the so-called ``intrinsic" PPA of the spin axis. Based on the observed RM$_{\rm obs}=-180.07\pm1.35$ on MJD 59456, we calculated the intrinsic position angle as $\psi_{0}$(intrinsic) = $\psi_{0}-$RM$\times\lambda^2$ = $89\fdg90\pm4\fdg60$, where $\lambda$ represents the observing wavelength. The degrees of linear and circular polarization for PSR~J0002+6216 are 78(4)$\%$ and 9(1)$\%$, respectively. 

%%%%%%%%%% figure PPA %%%%%%%%%%
%%%%%%%%%% figure PPA %%%%%%%%%%

%%%%%%%%%%%%%%%%%%%%%%%%%%%%%%%%%%%%%%%%%%%%%%%%%%
 \subsection{Spin-velocity relationship} \label{sec:Ali}
%%%%%%%%%%%%%%%%%%%%%%%%%%%%%%%%%%%%%%%%%%%%%%%%%%
In Paper I, one of the key pieces of evidence supporting the association between PSR~J0002+6216 and CTB~1 is that the tail of the pulsar's bow-shock wind nebula, located at a position angle of 113$^\circ$, points back towards the geometric center of CTB~1. From Table~\ref{tab:PM}, the position angle of the detected proper motion, as determined by both VLBI and timing, is consistent with the direction of the tail. This supports the associations between PSR~J0002+6216 and the bow-shock wind nebula, as well as between PSR~J0002+6216 and CTB~1. As discussed in Paper II, based on the proper motion measured through VLBI and the angular offset between PSR~J0002+6216 and the geometric center of CTB~1, it's concluded that PSR~J0002+6216 is a young pulsar with kinematic age of about 47~kyr. 

More recently, from the detection of the first 3D spin-velocity alignment in PSR~J0538+2817, \citet{2022ApJ...926....9J} proposed a novel model to explain pulsar spin-velocity alignment. This model predicts that young pulsars with high velocities tend to have smaller spin-velocity angle. For the young pulsar PSR~J0002+6216, given the higher precision of the proper motion measured from VLBI, we will utilize the VLBI results for the subsequent analysis of the spin-velocity relationship. From FAST polarization results and the VLBI proper motion measurements, we found that the 2D spin-velocity angle is $22\fdg96\pm4\fdg67$, which reveals a 2D spin-velocity misalignment. 

\begin{figure}[!ht]
\center
 \includegraphics[width=8.5 cm]{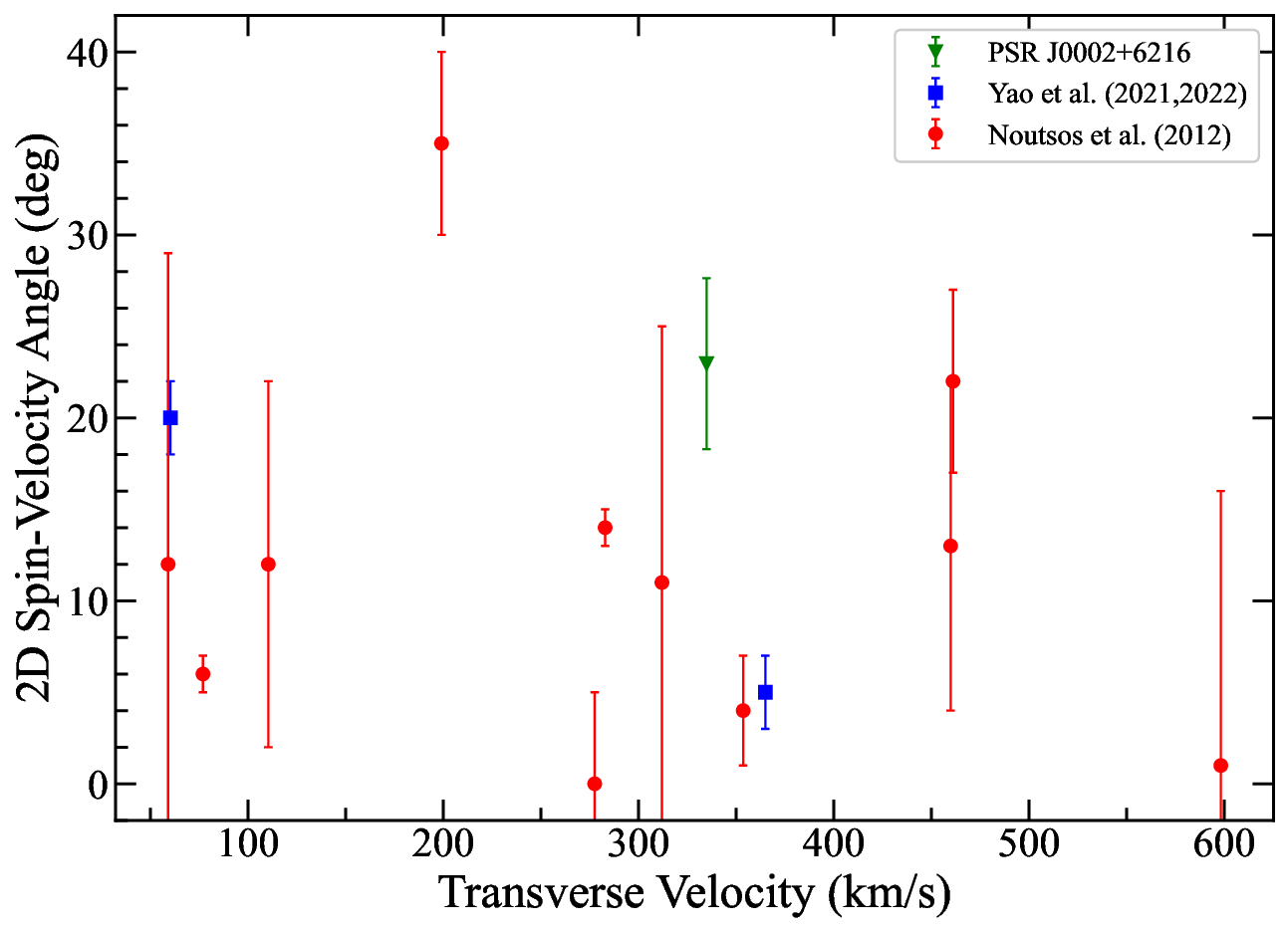}
    \caption{The 2D spin-velocity misalignment angle as a function of pulsar transverse velocity. These pulsars all have characteristic age less than $10^6$~yr and independent distance measurements. Green: PSR~J0002+6216; Blue: PSRs~J0538+2817 and B0656+14 from \citet{2021NatAs...5..788Y, 2022ApJ...939...75Y}; Red: pulsars from Table~1 of \citet{2012MNRAS.423.2736N}.}  \label{fig:spin-v}
\end{figure}

As depicted in Figure~\ref{fig:spin-v}, we have included the measurement of PSR~J0002+6216 in Figure 10 from \cite{2022ApJ...939...75Y}, to test the \citet{2022ApJ...926....9J} model, i.e., that spin-velocity correlation is higher for fast-moving pulsars than for slow-moving pulsars. The 3D spin axis for PSRs~J0538+2817, B0656+14 and J0002+6216 were derived from FAST observations, while the remaining datasets were sourced from Table~1 of \citet{2012MNRAS.423.2736N}. Although PSR~J0002+6216 has a similar velocity to PSR~J0538+2817, it exhibits a larger 2D spin-velocity angle. From Figure~\ref{fig:spin-v}, we found that the present data are insufficient to reliably verify or refute this model. Therefore, we require more high-sensitivity observations to confirm this. Future studies will benefit from polarization measurements of additional pulsars using the high-sensitivity observations of FAST core array \citep{2024AstTI...1...84J} and the ultra-wide band (UWB) observations of Qitai Radio Telescope (QTT) \citep{2023SCPMA..6689512W}, which will be crucial for examining the relationship between the spin and velocity of young pulsars.

%%%%%%%%%%%%%%%%%%%%%%%%%%%%%%%%%%%%%%%%%%%%%%%%%%
\section{Summary and Conclusions}\label{sec:4}
%%%%%%%%%%%%%%%%%%%%%%%%%%%%%%%%%%%%%%%%%%%%%%%%%%
In this paper, we conducted a timing and polarization study of the young pulsar PSR~J0002+6216 by combining observations from Fermi-LAT and FAST. Our timing analysis of PSR~J0002+6216 provided the proper motion and revealed two glitches for the first time. By comparing the timing results obtained from different software packages, we found that {\tt TEMPO2} and {\tt PINT} provided identical position measurements. From the positions provided by {\tt TEMPO2} and {\tt PINT}, we determined the total proper motion of PSR~J0002+6216 to be $\mu$$_{\rm tot}= 39.05\pm15.79$~mas~yr$^{-1}$, which aligns well with the VLBI measurement of $\mu$$_{\rm tot}=35.30 \pm 0.60$~mas~yr$^{-1}$, consistent within 0.24$\sigma$. Thus, through timing analysis, we confirmed that PSR~J0002+6216 is not a high-velocity pulsar. However, both measurements differ significantly from the proper motion values presented in Paper I. This discrepancy may be due to insufficient timing precision, but it could also be related to the fact that young pulsars are more prone to glitches. 

Both the Fermi-LAT and FAST data each showed a single glitch. The first, small glitch occurred on MJD~58850(17), with values of $\Delta \nu/\nu \sim  2.2(3)\times10^{-9}$ and $\Delta \dot{\nu}/\dot{\nu} \sim 1.0(3)\times10^{-3}$, identified through timing analysis of the Fermi-LAT data. More recently, we discovered that PSR~J0002+6216 experienced a spin frequency jump of $21243(7)\times10^{-9}$~Hz on MJD~60421(6) in the FAST timing data, marking its second glitch. The parameters for this large glitch are $\Delta \nu/\nu \sim  2450.7(8)\times10^{-9}$ and $\Delta \dot{\nu}/\dot{\nu} \sim 125(8)\times10^{-3}$. For the second glitch, we observed a distinct exponential recovery process in the post-glitch evolution of $\nu$, with $Q= 0.0090(3)$ and $\tau_{\rm d} = 45(3)$ days.

Based on observations with FAST, we measured the RM and the 3D spin axis of PSR~J0002+6216 for the first time and updated its DM. In FAST data we analyzed, we did not detect any significant variations in DM or RM. At a frequency of 1375~MHz, the linear polarization of PSR~J0002+6216 is approximately consistent with the total intensity, and the observed PPA can be fitted using the RVM. By combining the measured RM with the RVM-fit results, we determined that the projected orientation of the pulsar spin axis on the plane of the sky is $\psi_{0}(\rm intrinsic)=89\fdg9\pm4\fdg6$. When we compared this with the proper motion position angle presented in Paper II, we found a misalignment of approximately 23$^{\circ}$ between the pulsar's spin and velocity angles. Currently, pulsars with 2D spin-velocity angle measurements do not exhibit a trend that supports the \citet{2022ApJ...926....9J} model, which suggests that the higher the pulsar's velocity, the smaller the spin-velocity angle. However, we hope that with more high-precision observational data in the future, we will be able to further test models related to pulsar birth. 

%%%%%%%%%%%%%%%%%%%%%%%%%%%%%%%%%%%%%%%%%%%%%%%%%%
\section{Acknowledgements} \label{sec:5}
%%%%%%%%%%%%%%%%%%%%%%%%%%%%%%%%%%%%%%%%%%%%%%%%%%
This work is supported by the National Science Foundation of Xinjiang Uygur Autonomous Region (2022D01D85), the National Natural Science Foundation of China grant (No. 12288102), the CAS Project for Young Scientists in Basic Research (YSBR-063), the Chinese Academy of Sciences (CAS) “Light of West China” Program (Nos. 2022-XBQNXZ-015 and xbzg-zdsys-202410), the Major Science and Technology Program of Xinjiang Uygur Autonomous Region (No. 2022A03013-2) and the Tianshan talents program (2023TSYCTD0013). J.M.Y. is supported by the Tianchi Talent project. S.J.D. is supported by Guizhou Provincial Science and Technology Foundation (No. ZK[2022]304), the Major Science and Technology Program of Xinjiang Uygur Autonomous Region (No. 2022A03013-4), the Scientific Research Project of the Guizhou Provincial Education (No. KY[2022]132), and the Foundation of Education Bureau of Guizhou Province, China (Grant No. KY (2020) 003).

We thank the Fermi-LAT data archive for providing publicly accessible data used in this study. This study utilized data from the Five-hundred-meter Aperture Spherical radio Telescope (FAST), a large-scale national scientific facility in China, built and operated by the National Astronomical Observatories of the Chinese Academy of Sciences. We thank Kuo Liu, Lin Wang, and Shaungqiang Wang for valuable discussions. We would like to express our gratitude to all those who contributed to this research.

\bibliography{main}{}
\bibliographystyle{aasjournal}

\end{document}